\documentclass[prb,twocolumn,showpacs,amsmath,amssymb,a4paper]{revtex4}
\pdfoutput=1 

\usepackage{graphicx}
\usepackage{dcolumn}
\usepackage{bm}
\usepackage{hyperref}

\begin{document}

\title{Ultrafast time-resolved spectroscopy of 1D metal-dielectric photonic crystals}

\author{T. Ergin}

\author{T. H\"{o}ner zu Siederdissen }

\author{H. Giessen}

\author{M. Lippitz }
\email{m.lippitz@physik.uni-stuttgart.de}
\affiliation{4. Physikalisches Institut, Universit\"{a}t Stuttgart, 
Pfaffenwaldring 57, 70550 Stuttgart, Germany}
\affiliation{Max-Planck-Institut f\"{u}r Festk\"{o}rperforschung,
Heisenbergstr. 1, 70569 Stuttgart, Germany}

\date{\today}

\begin{abstract}
We study the all-optical switching behavior of one-dimensional metal-dielectric
photonic crystals  due to the nonlinearity of the free metal electrons. A
polychromatic pump-probe setup is used to determine the wavelength and pump
intensity dependence of the ultrafast transmission suppression as well as the
dynamics of the process on a subpicosecond timescale. We find ultrafast
(sub-picosecond) as well as a slow (millisecond) behavior. We present a model of
the ultrafast dynamics and nonlinear response which can fit the measured data
well and allows us to separate the thermal and the electronic response of the
system. \
\end{abstract}

\maketitle

\section{ Introduction}

In the late 1980's, the field of photonic crystals emerged
\cite{yablonovitch87,john87,joannopoulos95}. SInce then It has  rapidly evolved
and grown more interesting to a wide community. Photonic crystals show
remarkable properties due to their band structure, which lead to the realization
of a number of applications, such as bending of light around 120$^\circ$ corners
\cite{tokushima00}, extremely slow light in a medium \cite{gersen05} and
negative refraction while keeping the index of refraction positive
\cite{cubukcu03}.

Another promising field of application is the combination of photonic and
electronic devices, or even the replacement of electronics by photonics or
plasmonics \cite{macdonald09}. Such all-optical circuits have several
advantages over conventional electronics, such as reduced size, high repetition
rates, and enhanced speed of operation \cite{scalora83}.

A key component for such circuits is an all-optical switch. One-dimensional
metal-dielectric photonic crystals (1DMDPCs) might fulfill this purpose, as
small pump-induced variations in the dielectric properties can lead to strong
transmission changes. In this article we study experimentally the suitability of
1DMDPCs for ultrafast switching  in the subpicosecond regime, determine the
detailed wavelength and intensity dependences  and give a theoretical model,
which agrees well with our experimental data. 

A 1DMDPC is a structure which exhibits a periodic change of the index of
refraction in one dimension. Alternating layers of metal and dielectric are
deposited with a subwavelength thickness. This gives rise to a photonic
bandstructure with passbands and stopbands in the transmission spectrum. A large
amount of metal can thus be accumulated in these structures while keeping them
transparent \cite{scalora98}. This unique feature allows to exploit the
nonlinear properties of the constituting materials \cite{lepeshkin04}. Various
properties and possible applications have been proposed based on numerical
calculations and experiments, for example optical limiting, switching and
selective shielding \cite{scalora94,scalora98,scalora98_2,lee06_vandriel}.
However, the time constant of the nonlinear respose of such structures was not
measured. While previous work made use of the strong nonlinearity of the bound
metal electrons, we explore the possibility of using the response of the free
electrons. That approach has the advantage that no material resonance is
involved and thus the whole device could be tailored over a much larger spectral
range.

\section{Experimental techniques}

We use a polychromatic pump-probe setup to investigate the 1DMDPC. The laser
system consists of a Ti:Sa oscillator, which is pumped with 10~W at 532~nm. This
gives rise to 1.5~W average power and a pulse length of 200~fs at 825~nm. The
repetition rate of the modelocked laser system is 76~MHz. We split the beam into
a pump and a probe arm. The pump arm length is tunable mechanically, which
allows  high temporal resolution of our setup, limited only by the pulse
length. We modulate the pump beam with an acousto-optical modulator (AOM) at
100~kHz for lock-in detection. In a second beampath, a tapered fiber \cite{teipel03} is used to generate a
white light continuum from which the probe pulse is shaped in the time as well
as in the frequency domain. This allows us to choose the probe pulse central
wavelength and width while still keeping the pulse duration at 600~fs. The
pulses are overlapped in a polarizing beamsplitter cube and focused collinearly
on the sample. Behind the sample the pump beam is filtered out
spectrally,  and a balanced receiver detects the transmitted probe power.

The samples are prepared using an electron beam evaporation technique.
Alternating layers of Ag and MgF$_2$ are evaporated on a quartz substrate at
a pressure of 10$^{-7}$ mbar. The metal is evaporated with a rate of
20$\mathring{A}$/s to avoid clustering and to achieve smooth films, since otherwise silver
tends to cluster when applied at such small thicknesses
\cite{johnsonandchristy72,holland56}. The smoothness of the film, on the other
hand, is critical for consistent measurements.  For reasons that become obvious
below, we refer to a
layer series of metal--dielectric--metal as a cavity. Two samples were used in
this work: a single cavity sample (30~nm~Ag
and 225~nm~MgF$_2$) and a double cavity sample (20~nm~Ag and 215~nm~MgF$_2$).
The first has a narrow transmission resonance at 720~nm, the latter at 705 and
810~nm.

\section{Linear transmission properties}

\begin{figure}
	\centering
		\includegraphics[width=80mm]{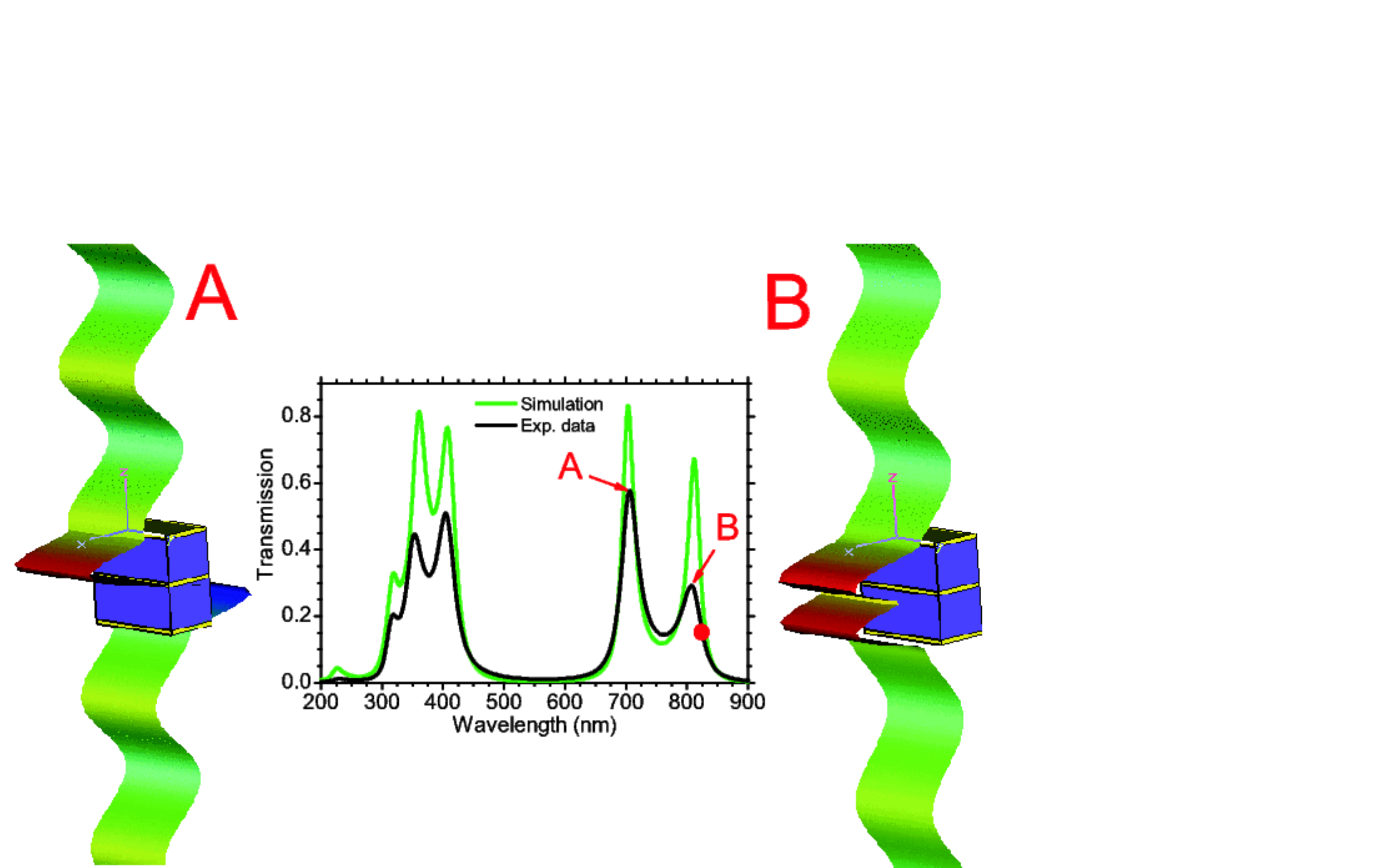}
	\caption{\textit{center:} Measured linear transmission spectrum of the
'double cavity' 1DMDPC. Passbands with multiple transmission peaks occur, marked
\textbf{A} and \textbf{B}. The simulated spectrum was calculated using the transfer matrix method. The red dot indicates the fixed pump wavelength used in
this paper. \textit{left and right:} The electric field distributions at the
wavelengths marked \textbf{A} and \textbf{B} superimposed on the structure (Ag:
 yellow, MgF$_2$: purple). The field forms standing waves, with a symmetric
(\textbf{B}) and anti-symmetric (\textbf{A}) mode and nodes only at the metal
layers. The transmission peaks at smaller wavelengths correspond to higher
order modes with an additional node in the center of each MgF$_2$ layer.}  
	\label{fig:cavity}
\end{figure}

In a first step, it is instructive to look at the linear transmission spectra of
1DMDPCs and their bandstructure. Figure~\ref{fig:cavity} shows a measured
transmission spectrum of the single cavity sample. Calculations using the
transfer matrix method \cite{pedrotti07_engl,macleod01} reproduce the features
very well. Each passband (around 700--850~nm and around
300--450~nm) is approximately 150~nm wide and consists of two peaks. The small
third peak in the passband at 310~nm is due to the interband transition of
silver. To understand these features, we consider the Bragg condition
\cite{husakou07,huebner99}. The metal layers form Bragg planes, which are separated by 
dielectric layers of  thickness $d$. This leads to standing waves inside the
structure, if the condition $n\cdot d=k\cdot\lambda_0/2$ is fulfilled, where
$n$ is the refractive index, $\lambda_0$ is the wavelength in vacuum and $k$ is
an integer.  However, the Bragg model is only a first approximation, as according to it 
 the long wavelength passband should occur at 675~nm and have only one peak.

We can also regard our system as coupled cavities. Each layer series
of metal--dielectric--metal forms a cavity. Consecutive cavities are coupled by
the electric field leaking through the joining metal layer. A series of $N$
cavities can be described by a series of $N$ coupled oscillators which exhibit
$N$ eigenmodes. Figure~\ref{fig:cavity} shows the electric field distribution
for the symmetric and anti-symmetric eigenmode of the lowest-order passband,
calculated with a FDTD method. The second-order modes (around 300--450 nm) have
an additional node of the field in the center of each cavity.

The  separation between two eigenfrequencies, i.e., two transmission peaks, is given
through the coupling strength of the oscillators, i.e., the thickness of the
silver layer. A thinner metal layer  corresponds to a stronger coupling of the
cavity system, since the evanescent light field decays exponentially inside the
silver. The  transmission is lower for the symmetric
fundamental mode (marked B), as the electric field undergoes a $\pi$ phase
jump at the silver layer, which leads to a higher field inside the metal and
therefore to a greater loss.

\section{Thermal effects}
\label{Thermal_effects}

\begin{figure}
	\centering
		\includegraphics[width=70mm]{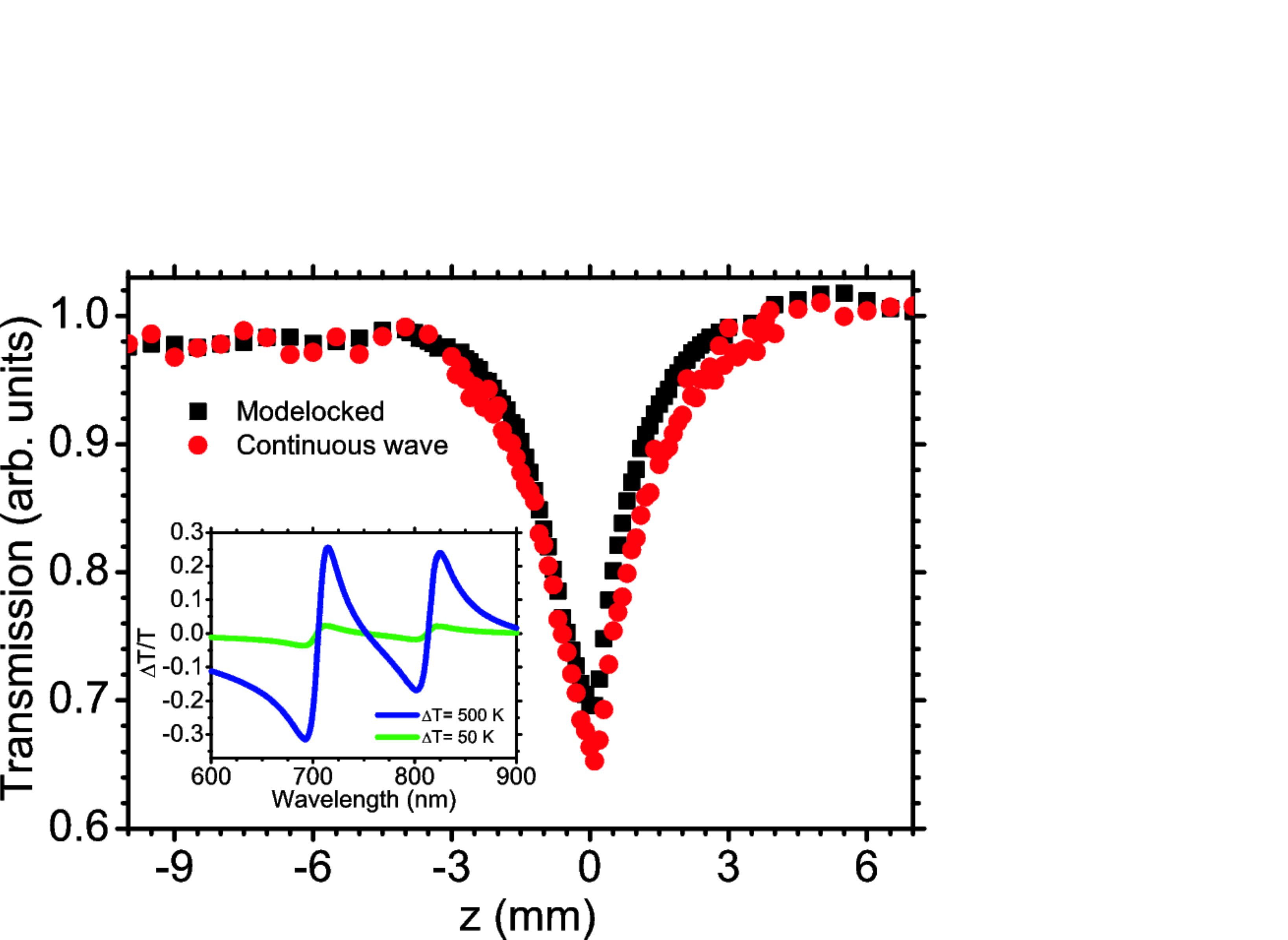}
	\caption{Open aperture z-scan measurement of the double cavity sample at
a pump wavelength of 825~nm (see
red dot in Fig.~\ref{fig:cavity}).  The sample was measured with a continuous
wave and a pulsed laser beam with the same average power (20~mW), but peak
powers differing by about five orders of magnitude. The inset shows the calculated
differential transmission due to thermal effects. The assumed temperature modulation is
$\Delta T=50~K$ and $500~K$ starting from room temperature.}
	\label{fig:slow}
\end{figure}

In contrast to experiments at a low laser repetition rate \cite{lepeshkin04}, our
repetition rate of 76~MHz leads to a larger power absorbed in the sample
at even moderate pulse energies. The temperature rise can reach several hundreds
of Kelvins locally, and thermal effects have to be considered. Any change in the
optical properties due to temperature will lead to a cavity detuning and thus to
a change in transmission which could mislead the interpretation of experiments
that do not contain  temporal information on the picosecond timescale, such
as z-scan experiments\cite{lepeshkin04,rotenberg07}. A z-scan \cite{sheik-bahae90} is a single beam
experiment, where the sample is moved through the focus along the optical axis
(z-axis) of the beam. The transmission is measured depending on the z-position
of the sample, i.e., as function of the laser intensity, as the focus size varies. Such measurements are used to determine the nonlinear absorption coefficient and
the nonlinear refractive index. Our laser system allows us to switch between
mode-locked and continuous wave mode, changing thus the peak power by about five
orders of magnitude without changing the average impinging power. Figure~\ref{fig:slow}
depicts the results of  an open aperture z-scan measurement of the double cavity
sample at a pump wavelength of 825~nm and a focal length of 50~mm. In our case,
the change in transmission is not due to effects that are nonlinear in the
optical field strength, but is only related to the deposited energy in the
system and thus the heatload. The thermal effect has the expected slow rise and
decay times (230 ms and 1570 ms, respectively).

A thermal response of the 1DMDPC can have several sources: thermal expansion due
to the thermal expansion coefficient $\delta$, change of the index of refraction
due to the thermo-optic coefficient $\Gamma$, change of the index of refraction due to the nonlinear refractive index
$n_2$, and the change of the plasma frequency of the metal in an expanded layer.
The last two points are negligible, since the layer expansion and thermo-optic
effects are orders of
magnitude larger for the materials used in our experiment. The inset in
Fig.~\ref{fig:slow} shows the calculated
differential transmission for temperature variations $\Delta T$ of 50~K and
500~K starting from room temperature using the transfer matrix method. The
parameters \cite{weber03} for this calculation are $\delta_{\text{Ag}}=1.89\cdot
10^{-5}~K^{-1}$, $\delta_{\text{MgF$_2$}}=1\cdot10^{-5}~K^{-1}$, and
$\Gamma_{\text{MgF$_2$}}=2\cdot10^{-6}~K^{-1}$. The calculated effect reaches the same order as the
measured transmission change  of the z-scan.

\section{Time-resolved ultrafast experiments}

To avoid the ambiguity of z-scan measurements with respect to thermal effects,
we now turn to time-resolved pump-probe experiments on the picosecond timescale.
They will allow us not only to separate the influence of heat accumulation on a
millisecond timescale from ultrafast effects, but also to gain the time
constants of the  ultrafast response of the 1DMDPC.

\subsection{Ultrafast response of metals}
\label{Ultrafast response of metals}
The  ultrafast response of the 1DMDPC is governed by the response of the metal
layers, which can be separated into four phases that overlap in time
\cite{bigot00,fann92,perner97}:

- Pump photons excite some free electrons into higher levels far above the Fermi
level, which leads to a non-Fermi-Dirac distribution, since most of the
electrons are still 'cold'. Our pulses of 600 fs length can not resolve this
short phase \cite{delfatti00jun15}.

- In a few tens to hundreds of femtoseconds, the electron gas thermalizes via
electron-electron scattering to a Fermi-Dirac distribution. Electron gas
temperatures of
several hundreds to a few thousand Kelvins can be reached. As our experiment in
the near infrared probes the free electrons, this phase  leads to the
highest system response.

- The electron gas transfers its energy to the lattice via electron-phonon
coupling. Due to the large heat capacity of the lattice, a single laser pulse
increases the temperature by  only a few ten Kelvins.

- The coupled system cools down to the same temperature as outside the laser spot. The
accumulated absorbed energy together with the slow heat conduction away from the
laser spot leads to the temperature increase of some hundreds of Kelvins, as
discussed in the last section.

As soon as a thermal distribution of the excited electrons is reached, the
coupled electron-lattice system can be described by the two-temperature model
(TTM) \cite{anisimov74}. The temperatures of the
electron gas and the lattice ($T_e$ and $T_l$) are given by two coupled
differential equations:
\begin{align}
C_e\frac{\partial T_e}{\partial t} & =-g\left(T_e-T_l\right)
+\vec{\nabla}\kappa\vec{\nabla}T_e + S\\
C_l\frac{\partial T_l}{\partial t}&=g\left(T_e-T_l\right) \quad .
\end{align}
$C_e=\alpha T_e$ is the specific heat capacity of the electron gas, $\kappa$ is
the heat diffusion
coefficient, $S$ is the source term, $g$ describes the electron-phonon coupling,
and $C_l$ is the specific heat capacity of the lattice. The temperatures and the
source term are functions of space and time. To simplify the calculations, we
avoid a full spatial treatment of the system by neglecting the spatial
dependence and by replacing the exact heat diffusion term with an empiric
diffusion term including a fit parameter $d$, which couples the system to a
temperature bath at room temperature $T_{\text{room}}$. This accounts for the
heat transport out
of the system, which is hereby assumed to be governed by the electron gas. The
equations are then reduced to
\begin{align}
\label{eq:TTM}
	\frac{\partial T_e}{\partial t}&=-\frac{g\left(T_e-T_l\right)}{\alpha
T_e}-d\left(T_e- T_{\text{room}} \right)+\frac{S}{\alpha T_e}\\
	\frac{\partial T_l}{\partial t}&=\frac{g\left(T_e-T_l\right)}{C_l} \quad .
\end{align}
The parameters \cite{bigot00} for the simulations are $\alpha=66J/(m^3K^2)$,
$g=3\cdot10^{16}W/(m^3K)$, and $C_l=2.415\cdot10^6J/(m^3K)$. We chose $C_l$ to be
constant, since we measure at a temperature well above the Debye-temperature of
silver ($\approx 200~K$). The source term is assumed to be a Gaussian pulse (in
time) which is absorbed in a spot of 10~$\mu m$ diameter. For reasons of simplicity, we assume the whole pulse energy to be absorbed in these numerical calculations.

For the calculation of the transient transmission suppression we incorporate the
TTM into the transfer matrix and reference the work of Bigot \textit{et al.}
\cite{bigot00}. The electron gas and lattice temperature are
calculated using the TTM. To account for the change of the optical properties,
the dielectric function $\epsilon(\omega)$ of the metal layer is modified, starting from Johnson
and Christy values \cite{johnsonandchristy72}, with a Drude term for the free electrons that depends on the rise of the
electron gas temperature  $\Delta T_e$ :
\begin{align}
\epsilon(\omega, \Delta T_e) = \notag &\epsilon_{\text{J+C}}(\omega) -
\epsilon_{\text{Drude}}(\omega
,\Delta T_e = 0) \\&+ \epsilon_{\text{Drude}}(\omega, \Delta T_e) \quad.
\end{align}
The Drude model is made temperature dependent by letting the damping constant
$\gamma$ depend on  $\Delta T_e$. This is motivated as with increasing electron
gas temperature the Fermi distribution smears out and more and more
possibilities for electron--electron scattering occur. We set\cite{bigot00}
\begin{align}
\label{eq:gamma}
	\gamma(\Delta T_e)=\gamma_0 + \beta \cdot \Delta T_e \quad ,
\end{align}
where $\beta$ is a fit parameter in our calculations and $\gamma_0$ the room
temperature value of the damping constant. The modified Drude model reads thus
\begin{align}
\epsilon_{\text{Drude}}(\omega, \Delta T_e) = \epsilon_{\infty} - 
      \frac{\omega_p^2}{\omega \left( \omega + i \gamma \left(\Delta T_e \right)
\right)}  \quad ,
\end{align}
where $\omega_p$ is the plasma frequency.

\subsection{Temporal dynamics of the nonlinear optical response}

\begin{figure}
	\centering
		\includegraphics[width=70mm]{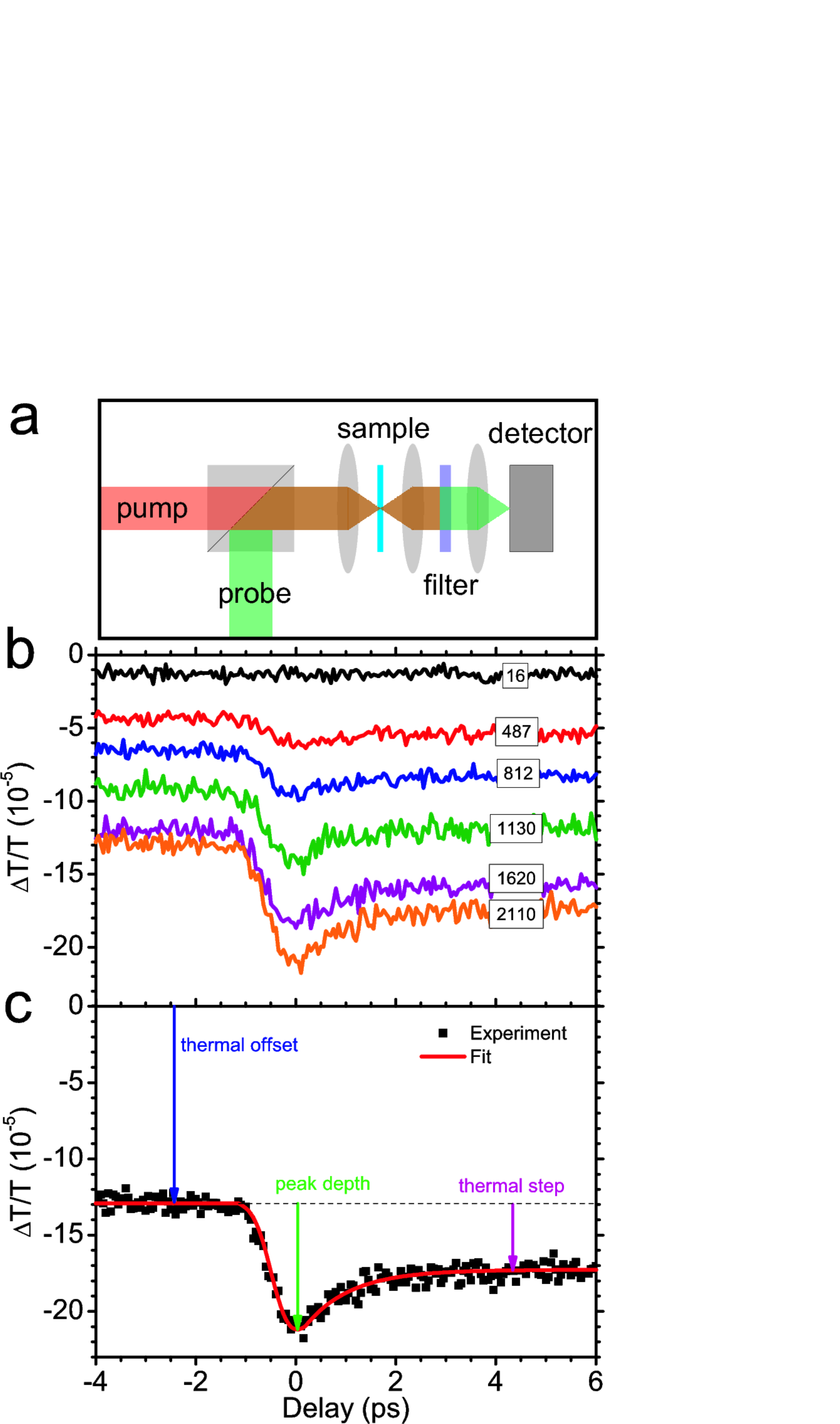}
	\caption{(a) Detection part of the experimental setup. (b) Pump-probe measurements
with varying pump intensity (given in MW/cm$^2$ as numbers in the graph) on the
single cavity sample. The absolute zero delay was set to the differential
transmission peak minimum. The sample was pumped at 825~nm and probed at 720~nm. (c) Example of
the model function (Eq.~ \ref{func:model}) fitted to one data set from (b).}
	\label{fig:intens}
\end{figure}

Our pump-probe setup (Fig.~\ref{fig:intens}a) allows us to investigate not only the amplitude but also
the temporal dynamics of the nonlinear optical effects in 1DMDPCs. A pump pulse
is absorbed, modifies the dielectric properties of the metal layers, and a probe
pulse interrogates the transmission of the photonic crystal after a variable
delay. Figure~\ref{fig:intens}b shows a series of such measurements on the
single cavity sample with varying intensity. The differential transmission is
not zero for negative delays (i.e., probe arrives before pump) because of the
thermal effect discussed above. This offset decreases with increasing AOM
modulation frequency, as less heat is accumulated in each AOM cycle. The probe
transmission drops at the arrival of the pump pulse and recovers on a
subpicosecond time scale. It does not recover fully to the start value, since
also the last pump pulse deposits additional heat in the focal area.

To extract quantitative data from these traces and similar ones from the double
cavity sample, we fitted a phenomenological model function $f(\tau)$ to the
data:
\begin{align}
  f(\tau) = A + \left[ B e^{- \tau / \tau_0} + C ( 1 - e^{- \tau / \tau_0} )
\right]
\Theta(\tau)  \label{func:model} \quad .
\end{align}
$A$ describes the amplitude of the thermal offset at negative delays. $B$ is the
amplitude of the response caused by the hot electron gas that decays
exponentially with a time constant $\tau_0$. The thermal effect of the pump
pulse is taken into account by a level $C$ that is exponentially approached with
the same time constant $\tau_0$. The last two processes start as soon as the pump
pulse arrives, i.e., the Heaviside step function $\Theta(\tau)$ switches from $0$
($\tau < 0$) to $1$
($\tau > 0$). To take into account the limited temporal resolution of our setup,
we convolute the model function $f(\tau)$ with a measured cross-correlation
function between pump and probe pulse before fitting it to the data. A fit to a
temporal trace at high pump power is shown in figure~\ref{fig:intens}c.

\subsection{Intensity dependence of the nonlinear optical response}

\begin{figure}
	\centering
		\includegraphics[width=70mm]{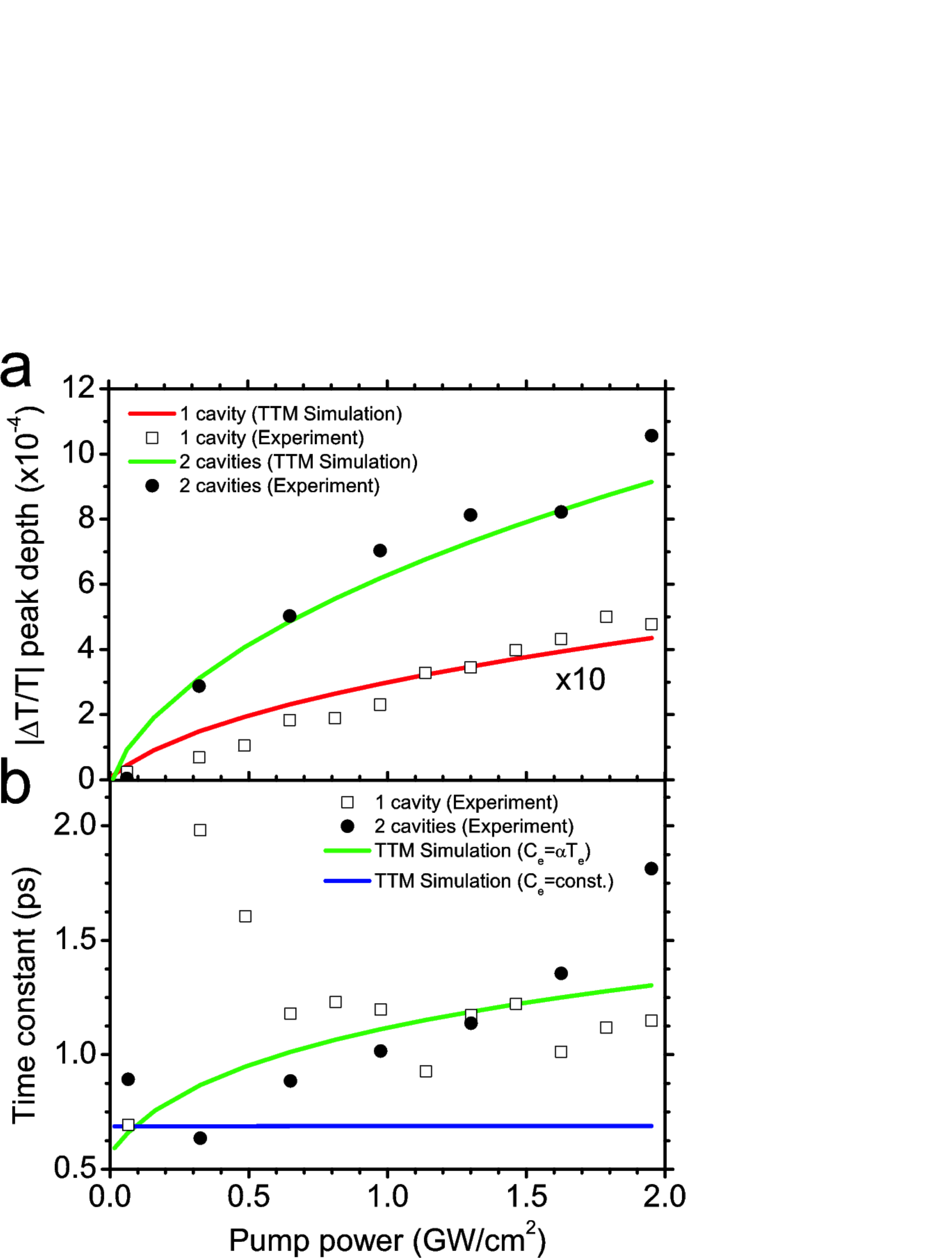}
	\caption{The TTM combined with a temperature-dependent Drude model describes with one parameter set pump-power dependence of both (a) the peak depth (amplitude $B$)  as well as (b) the recovery time constant $\tau_0$ of the ultrafast transmission suppression. The latter is  weakly dependent on the pump intensity due to the temperature dependence of the electron heat capacity.}
	\label{fig:time}
\end{figure}

The amplitude $B$ and the decay time $\tau_0$ of the nonlinear optical response
were recovered by fitting the phenomenological model function $f(\tau)$ to a set
of temporal traces with different pump intensities. Figure~\ref{fig:time} shows the data for
the single and the double cavity sample.  The absolute value  of
the nonlinear optical response $|B|$ seems to grow linear with pump intensity for both
samples, always maintaining its negative sign. The decay time $\tau_0$ is not
constant but also increases with pump intensity. Both effects can be described by a TTM that modifies the Drude damping in the metal's dielectric properties as
described in section~\ref{Ultrafast response of metals}. Only two fit parameters are used for both parts of Fig.~\ref{fig:time}: firstly the relation
between the increase in electron gas temperature $\Delta T_e$ and Drude damping
$\gamma$, described by $\beta$ in equation~(\ref{eq:gamma}), and secondly the empirical constant $d$
to describe the heat diffusion by electrons in equation~(\ref{eq:TTM}). 
The fit leads to $\beta = 6\cdot 10^{-8}~eV/K$ and $d=3.7\cdot 10^{11}~1/s$. The deviations at small pump intensities for the single cavity sample in figure~\ref{fig:time}b might occur due
to a lesser sample quality and to slightly different probe locations on the
sample, which lead to locally changed heat diffusion and therefore to changed
dynamics.

The temporal response in the 1DMDPCs is on the order of 1~ps.  The effect of rising time
constants with increasing pump intensity can be understood, since the heat
capacity of the electron gas is proportional to its temperature ($C_e=\alpha
T_e$). The lattice heat capacity is orders of magnitude larger, but higher pump
intensities lead to a smaller difference between electron gas and lattice heat
capacity. This leads to a slower equilibration (longer response time) and to a
slightly higher end temperature. When the electron gas heat capacity is assumed
to be constant, for example at its value at $T_e = 400$~K
($C_e=2.64\cdot10^4~J/(m^3K)$), then the time constant is independent of the
pump intensity (see blue line in figure~\ref{fig:time}b). This  confirms the temperature
dependent electron heat capacity as origin of the pump intensity
dependence of the decay time. The general trend of
increasing response times at higher pump intensities was also found in single
layers of metal \cite{ali91}. The values of the time constants were on the order
of 1-2~ps.

To compare the magnitude of the nonlinear optical response of the 1DMDPCs with
other materials, it is instructive to calculate the phenomenological nonlinear
absorption coefficient $\alpha_{\text{eff}}$. It describes the intensity
dependence of the absorption coefficient $\alpha$
\begin{align}
       \alpha(I) = \alpha_0 + \alpha_{\text{eff}} \, I .
\end{align}
As the absorption coefficient $\alpha$ is connected to the transmission $T$ via
the sample thickness $L$: $T = T_0 e^{-\alpha L}$, the nonlinear absorption
coefficient $\alpha_{\text{eff}}$ is just a measure for the slope of a linear fit
to the data in figure~\ref{fig:time}a. We obtain values of $5.2\cdot10^{-10}~cm/W$ and $7.4\cdot10^{-9}~cm/W$ for
the single and double cavity sample, respectively, using as sample thickness $L$ the sum of all metal and dielectric layers in the multilayer structure.

\subsection{Wavelength dependence of the nonlinear optical response}

\begin{figure}
	\centering
		\includegraphics[width=80mm]{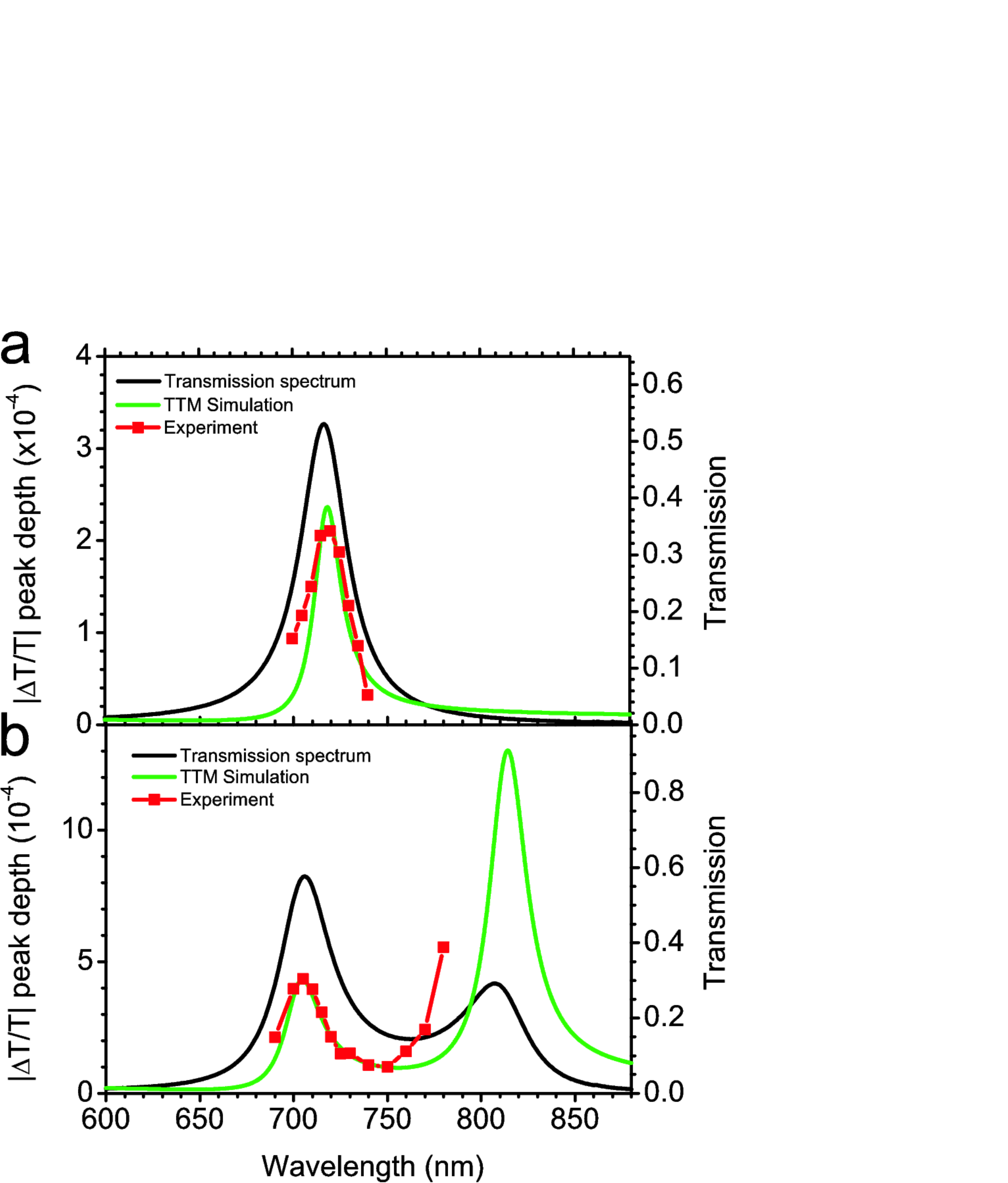}
	\caption{Spectral dependence of the transient transmission compared to the linear transmission spectrum and the model for  (a) the single cavity sample and (b) the double cavity sample. The pump power
(1.3~GW/cm$^2$) and pump wavelength (825~nm) was kept constant. The TTM
simulations give a good agreement.}
	\label{fig:wave_stacked}
\end{figure}

Figure~\ref{fig:wave_stacked}a shows the wavelength dependence of the ultrafast
transmission suppression in the single cavity sample. For this measurement, pump
power and pump wavelength were kept constant (1.3~GW/cm$^2$ at 825~nm). The
shape of the differential transmission data follows the linear transmission
spectrum, which means that the spectral response of an all-optical switch could
be tuned by adjusting the thickness of the metal and dielectric layers.

The double cavity sample shows a similar behavior
(Fig.~\ref{fig:wave_stacked}b). The TTM simulations predict a stronger response
at the 810~nm transmission peak compared to the peak at 705~nm. The same trend
is found in the experiment. However, the  probe spectral range is limited by the
pump filter stopband starting at 780~nm. The measurement shows a steeper rise
than the simulations at 775~nm. This might be due to the roughness and
inhomogeneity of the silver films. The position of the 810~nm peak is dependent
on the silver layer thickness, and the position of the 705~nm peak is relatively
insensitive to thickness variations. The white light transmission spectrum is
measured with a large spot (4~mm in diameter), which leads  averages over the
roughness. The pump-probe measurements, on the other hand, were performed with a
spot size of roughly 10~$\mu m$. If a region with a slightly larger silver
thickness was measured, the result would be a shift of the 810~nm peak into the
blue, which would explain the rise at smaller wavelengths.  The TTM parameters for
these simulations are $\beta=1.1\cdot10^{-8}~eV/K$ and
$\beta=3.5\cdot10^{-8}~eV/K$ for the single and the double cavity sample,
respectively.

\section{Conclusion}

We have shown that one-dimensional metal-dielectric photonic crystals respond on
two timescales to a near-infrared pump pulse train  at a Megahertz repetition
rate: on a timescale of hundreds of milliseconds thermal effects lead to  a
large transmission change of 30--40\%. Simulations based on the transfer matrix
were presented to explain the measurements. On a picosecond timescale, the
ultrafast response of the metal's free electrons dominates. Its magnitude is in
the order of $10^{-5}$ in a single cavity system and $10^{-3}$ in a double
cavity system. The two-temperature model and a temperature dependent Drude model
explain both the pump intensity and the probe wavelength dependence of the
effect. Using the nonlinear response of the free metal electrons makes it
possible to tune the spectral response of the metal-dielectric photonic crystal
by its structure, i.e., without changing the material. However, that flexibility
is accompanied by a reduced nonlinearity compared to material resonances
\cite{rotenberg07,lee06_vandriel}.

\acknowledgments

We thank Dr.~S.~Linden (University of Karlsruhe) for the preparation of a
sample. We acknowledge financial support by  DFG (FOR 557 and 730) and BMBF (13N9144 and 13N10146).




\end{document}